
\hfuzz=10pt
\def\refmark#1{[#1]}  

\def\OMIT#1{}
\def\dt{{\rm d}t}
\def\dx{{\rm d}x}
\def\dtau{{\rm d}\tau}
\def\dk{{\rm d}k}
\def\domega{{\rm d}\omega}
\def\dsigma{{\rm d}\sigma}

\date={October 1993}
\pubtype={}
\Pubnum{\vbox{ \hbox{BROWN-HET-925} \hbox{hep-th/9310190} }}

\titlepage

\title{\seventeenbf
{Time-Dependent Backgrounds in 2D String Theory
and the
$S$-matrix Generating Functional}
\footnote{\star}{Research supported in part by
   the U.S.~Department of Energy under contract DE-AC02-76-ER03130.
   E-mail: {\tt   lee@het.brown.edu}
}
}

\author{  Julian Lee  }
\address{ Department of Physics, Brown University, Providence,
          RI 02912 }

\abstract

We study  time-dependent tachyon backgrounds of the two-dimensional
string collective field theory using the
formalism of the $S$-matrix generating functional. In the process we
clarify the connection between two ways of calculating the $S$-matrix,
the one using the Feynman rule and the other using the classical solution to
the nonlinear equation of motion.
We develop the formalism for general backgrounds and apply it
to the gravitational sine-Gordon model in detail.
We reproduce the conformal field theory calculation which was based
on expanding around the static $c=1$ theory.
Furthermore, we prove that the tree-level partition function of
this model shows the scaling behavior
corresponding to $c=0$ model in the limit
$p \to 0$ of sine-Gordon `momentum' $p$.

\endpage

{\bf \chapter{Introduction}}

In recent years we have seen a detailed study of 2-D string theory. Most of the
investigations and their results concern a particular background described in
conformal field theory by the $c=1$ matter + Liouville fields. This background
representing tachyon condensation was completely solved by a particular
(inverted oscillator) matrix model and the associated collective field theory
\REFS\mil{D.~J.~Gross and N.Miljkovi\'c \journal Phys.Lett.& B238 (1990) 217;
\nextline
E.Brezin,V.Kazakov,and Al.B.Zamolodchikov \journal Nucl.Phys. & B338 (1990)
673 ;\nextline
P.Ginsparg and J.Zinn-Justin \journal Phys.Lett. & 240B (1990),333; \nextline
G.Parisi \journal Phys.Lett. & 238B (1990) 209 ;
 D.J.Gross and I.R.Klebanov \journal Nucl.Phys.& B352 (1991) 671 }
\REFSCON\das{ S.R.Das and A.Jevicki \journal Mod.Phys.Lett.& A5 (1990) 1639}
\REFSCON\sakita{ A.Jevicki and B.Sakita \journal Nucl.Phys. & B165 (1980) 511
}
\REFSCON\kr{ K.Demeterfi, A.Jevicki and J.P.Rodrigues \journal Nucl.Phys.&
B362 (1991) 173;  K.Demeterfi, A.Jevicki and J.P.Rodrigues \journal
Nucl.Phys.&    B365 (1991) 499 }
\REFSCON\polch{ J.Polchinski \journal Nucl.Phys. & B362 (1991) 125}
\REFSCON\moreeq{G.Moore and R.Plesser \journal Phys.Rev. & D46 (1992) 1730 }
\REFSCON\ex{G.Moore and R.Plesser, and S.Ramgoolam \journal Nucl.Phys. &
B377 (1992) 143 ; G.Moore \journal Nucl.Phys. & B368 (1992) 557 }
\refsend .

One is in general interested in studying the theory in other classical
backgrounds. One such example is the sine-Gordon background investigated
recently in refs.
\REFS\moresine{G.Moore, YCTP-P1-92, hep-th/9203061 }
\REFSCON\hsu{E.Hsu and D.Kutasov,  {\sl Nucl.\ Phys.} {\bf B396} (1993) 693,
hep-th/9212023 }
\REFSCON\sch{ C.Schmidhuber, CALT-68-1891,  hep-th/9308120, Presented at
International Conference on Strings 93; C.Schmidhuber \journal Nucl.Phys. &
B404 (1993) 342 }
\refsend.
  Here the Liouville term is replaced by a general
tachyon (vertex) perturbation

$$ \Delta S = \int \lambda V_p(X,\phi) d^2z $$
and the phase diagram of the perturbed theory is found using the known
correlation functions of the c=1 model.

It is interesting to study these and other more general backgrounds in terms of
(string) field theory. For two dimensional strings all features of tachyons are
given by the collective field theory, which is induced from the matrix model.
There the c=1 background is given by a static solution of the scalar field
$\phi_0 = \sqrt{ \mu - x^2 }$. One has more general time dependent solutions
$\alpha_{cl}(t,x)$ which can be given in parametric form. The question that one
can address is specifying a particular field theory solution which gives the
sine-Gordon
background of string theory. This is a question of describing different
backgrounds in the framework of a single (string) field theory. It is also a
question of understanding further the nontrivial correspondence between matrix
models and the usual worldsheet conformal field theory.

We address these questions in this paper. Using the general classical solution
of collective equations we identify particular conformal field theory
backgrounds. Since the conformal description represents an $S$-matrix
generating
function the identification of backgrounds and correspondence with the
conformal field theory is done at the asymptotic level of linearized fields. It
is there that a clear identification applies and the full theory is then
perturbed around the corresponding nonlinear classical solution. We describe
this in detail for the particular case of a sine-Gordon background.

  The content of the paper goes as follows. First we discuss the $S$-matrix
generating functional for the collective field theory. It is given generally in
terms of the classical action and the classical solution with arbitrary initial
(asymptotic) conditions. An explicit equation between the linearized field and
the parameter describing the general classical solution is given. In the
process,
 we make a connection between two seemingly different ways of
evaluating the $S$-matrix, namely the direct use of the Feynman rule
\refmark{\kr} and
the method where one examines the asymptotic behavior of the classical
solution to the equation of motion \refmark{\polch,\moreeq}.
 In section 3
we apply the formulas derived to specify the nontrivial classical configuration
corresponding to the sine-Gordon background. The nontrivial scaling
corresponding to the $c=0$ model is then seen to follow by evaluating the
action for this solution.

{\bf \chapter{$S$-matrix Generating Functional in Collective Field Theory}}

The $S$-matrix generating functional is powerful for calculating scattering
amplitudes \Ref\fad{I.Ya.Aref'eva, A.A.Slavnov, and L.D.Faddeev \journal
Theor.Mat.Fiz.
\nextline & Vol.21,No.3 (1974) 311;
C.K.Lee and A.Jevicki \journal Phys.Rev & D37 (1988) 1485 }.
It was applied to the $\beta$-function equations of string theory in
ref.~\Ref\frid{ B.E.Fridling and A.Jevicki
\journal Phys.Lett. & 174B (1986) 75  }.
In this section we review the
formalism and apply it to collective field theory.

Consider first a field theory described by the action
$$
   I(\varphi) = \int \dx\, \dt\, \left( {1 \over 2 }
   \varphi(x,t) \ \hat O
   \ \varphi(x,t) + {\cal L}_{int} (\varphi)\right),
   \eqn\ACTION
$$
where~$ \hat O $ is a differential operator with
no explicit dependence on~$x$ or~$t$
and~$ {\cal L}_{int} $ is treated as a perturbation.
The equation of motion is given by
$$
   \hat O \ \varphi = -{\delta {\cal L}_{int}(\varphi) \over \delta \varphi},
   \eqn\EOM
$$
which can be integrated to yield the classical field
$$ \varphi_{\it cl}(x,t) = - \int \dx'\, \dt'\,
\Delta(t-t',x-x')  {\delta {\cal L}_{int}(\varphi_{\it
cl}(x', t')) \over \delta \varphi_{\it cl}(x', t')} +
\tilde \varphi(x,t).\eqno\eq $$
The propagator $ \Delta(t-t',x-x')$ is inverse of $ \hat O $
  and $\tilde \varphi $ is a linear field  satisfying
$$ \hat O \ \tilde \varphi = 0. $$
Since $\Delta$ is not unique,  neither is $\tilde \varphi $.
Among the many possible choices for the propagator,
the useful ones are retarded
propagator $\Delta_R(t-t', x-x')$ and advanced propagator
$\Delta_A(t-t', x-x')$ which satisfies
$$
   \eqalign{\Delta_R (t,x-x') & = 0 \;\;\;\; {\rm for} \; t<0  \cr
            \Delta_A (t,x-x') & = 0 \;\;\;\; {\rm for} \; t>0  \cr }
   \eqn\ra
$$
and the Feynman propagator $\Delta_F$ which is defined by
$$
   \tilde \Delta_F (\omega, k) \equiv \theta(\omega)
   \tilde \Delta_A (\omega, k) + \theta(-\omega)
   \tilde \Delta_R (\omega, k),
   \eqn\feyn
$$
where $\tilde \Delta ( \omega, k)$ is the
Fourier transform of $\Delta (  t, \tau)$.

Let us denote $\tilde \varphi$ as $\varphi_0$ when $\Delta = \Delta_F $ .
Then
treating  $\varphi_{\it cl}$ as a functional of $\varphi_0$,
we can write down
the tree-level $S$-matrix generating
functional, derived in ref.~\refmark{\fad},
$$
   \eqalign{ S(\varphi_0) &=\int \! \! \! \int \dx \dt\,
   \left( {1 \over 2 }
   (\varphi_{\it cl}-\varphi_0 ) \ \hat O \
   (\varphi_{\it cl}-\varphi_0)
   + {\cal L}_{int} (\varphi)\right) \cr
   & = \sum_n {1 \over n!} S_n(k_1,k_2,\cdots , k_n)
   \, \varphi_0(k_1) \varphi_0(k_2)
   \cdots \varphi_0(k_n) \cr }
   \eqn\gene
$$
where $S_n$ is the $n$-point scattering amplitude calculated from the
perturbation theory and $\varphi_0(k)$ is the Fourier transform of
$\varphi_0(x)$.
As we show in section~3, the $S$-matrix generating functional in
collective field theory  corresponds to the partition
function in string theory
with general tachyon background.

Note that when we substitute $\varphi_{\it cl}(\varphi_0)$
into~$I(\varphi)$, it differs from $S(\varphi_0)$ by a
nontrivial boundary term,
$\varphi_0 \hat O \varphi_{\it cl}$.
(The piece $(\varphi_{\it cl}-\varphi_0) \hat O \varphi_0 $
vanishes by the equation
of motion $\hat O \varphi_0 = 0 $.)
This prescription for the surface term --- in which one does not freely
integrate~$\gene$ without accounting for the boundary term ---
gives the correct $S$-matrix\refmark{\fad}.

Using the identity
$$ \varphi_{cl} = -\Delta_F \ { \delta S(\varphi_0) \over \delta \varphi_0 }
  \eqn\?
$$
we get the simple relation between action $I(\varphi_{cl})$ and the
functional $S(\varphi_0)$,
$$
   \eqalign{ I(\varphi_{cl})
   &=  {1 \over 2} \varphi_0 \ \hat O \ \varphi_{cl}
   + S(\varphi_0)
  = -{1 \over 2} \varphi_0 { \delta S(\varphi_0) \over \delta \varphi_0 } +
   S(\varphi_0)
   \cr &
   = -{1 \over 2 }  \sum_n { n-2 \over n!} S_n(k_1,k_2,\cdots , k_n)
   \varphi_0(k_1) \varphi_0(k_2) \cdots \varphi_0(k_n) \cr }
   \eqno\eq
$$

This tells us that given an
exact solution to the classical equation of motion eq.~$\EOM$,
we construct the desired $S$-matrix generating
functional by substituting it into the action~$\ACTION$,
with careful redefinition of the quadratic term.
Now we turn to string theory.

The collective field theory describes
string field theory in~$1+1$ dimensions.
Its action is given by \refmark{\das, \sakita}
$$ I = \beta^2 \int \dx \dt\, \left[  {1 \over 2}   {  (
\partial_x^{-1}
\dot \varphi )^2  \over \varphi } - { \pi^2 \over 6 } \varphi^3 + {1
\over 2 } (x^2+\mu) \varphi \right] \eqno\eq $$
We assume $\mu > 0 $ for convenience.(Above barrier.) For $\mu < 0$, we can get
 the result by analytic continuation of $\mu$ . We can rescale
$$
   x  \to ( \beta g )^{-{1 \over 2}} x
   , \qquad
   \varphi \to ( \beta g)^{-{1 \over 2}} \varphi ,
   \eqn\?
$$
to get
$$
   I = {1 \over g^2} \int \dx \dt\, \left[  {1 \over 2}   {  (
   \partial_x^{-1}
   \dot \varphi )^2  \over \varphi } - { \pi^2 \over 6 } \varphi^3 + {1
   \over 2 } (x^2+1) \varphi \right] \eqno\eq
$$

To apply the formalism developed above we
first rewrite this action in terms of left- and right-moving
fields
$$
   p_\pm = g_s^2 \partial_x \pi_\varphi \pm \pi \varphi ,
   \eqn\?
$$
where $ \pi_\varphi $
is the canonical momentum conjugate to the field $\varphi(x,t)$.
The action eq.~$\ACTION$ becomes\refmark{\polch}
$$ I= { 1 \over 2\pi g^2 } \int \dx \dt\, \left[ -{1 \over 2}  p_+
\partial_x^{-1} \dot p_+ +{1 \over 2} p_-
\partial_x^{-1} \dot p_-  -  { p_+^3 \over 6 }  +  { p_-^3
\over 6 }  +  {1 \over 2 } ( x^2  +  1 )(p_+  -  p_-) \right]
\eqno\eq
$$

The equation of motion is
$ \partial_t p_\pm = x - p_\pm \partial_x p_\pm $
and its static solution is
$ p_\pm = \pm \sqrt { x^2 + 1 } $.
We may expand the field around this background,

$$ p_\pm= \pm \sqrt { x^2  + 1 } + g \eta_\pm \eqno \eq, $$
where $\eta$ denotes the time-dependent fluctuation
and the explicit factor of the coupling~$g$ denotes that it
vanishes with the interactions.
In terms of~$\eta$ the action is
$$
   \eqalign{
   I= { 1 \over  4 \pi } \int \dt \dx\, \bigg[ - \eta_+
   \partial_x^{-1}  \dot { \eta_+ }
   &
   + \eta_-
   \partial_x^{-1}  \dot { \eta_- } \> - \> ( \eta_+^2
   \, + \, \eta_-^2 ) \sqrt { x^2 \, +1 } \>
   \cr &
   - { { g \,
   \eta_+^3 } \over 3 } \, + \, { { g \, \eta_-^3 } \over 3 }
   \bigg] + S_0
   .\cr} \eqn\?
$$
where $S_0$ is the action for to the $c=1$ vacuum.
This piece scales like $ {1 \over g^2} \log \mu $ \refmark{\das}.
The additional $\log \mu$
dependence comes from the limits of space integration.
For convenience, let us drop this term from now on.
Taking $ x = \sinh \tau $
and defining $ \alpha _ \pm \equiv   \eta_\pm  \cosh \tau $
we get
$$
   I^\pm = \mp { 1 \over 4 \pi } \int \dt \dtau\, \left[
   \alpha_\pm \partial_\tau^{-1} \dot \alpha_\pm \>
   \pm \> \alpha_ \pm ^2 + { g \over 3  } \,
   { \alpha _ \pm ^3 \over \cosh ^2 \tau } \right]
   \eqn\?
$$
where
$ I = I^+ \, + \, I^- $.

Since $\alpha_+$ and $\alpha_-$ are decoupled, it is sufficient
to study one branch.
Let us take~$\alpha_+$ and drop the subscript for clarity.
The equation of motion is
$$ ( \partial_t +\partial_\tau ) \alpha =  -{g\over 2}
\partial_\tau \{ {\alpha^2\over \cosh^2 \tau} \} \eqno\eq $$

   We would like to express $\alpha$ in terms of
the linearized field $\alpha_0$, which is defined to satisfy
the free equation
$(\partial_t +\partial_\tau ) \alpha_0  = 0 $.
The two are connected through the integral relation
$$
   \eqalign{ \alpha_0(t,\tau) &\equiv \alpha(t,\tau) - \int
   \dt' \dtau' \Delta_F (t-t' , \tau - \tau' ) (
   \partial_{t'} + \partial _ { \tau '} ) \alpha (t',
   \tau') \cr
   &= \alpha (t,\tau )   +
   \int \dt' \dtau' \Delta_F (t-t' , \tau - \tau' )
   \, {g\over 2}\,  \partial_{\tau' } \left \{ { \alpha^2 (t',
   \tau ') \over \cosh^2 \tau } \right \} \cr }
   \eqn\lin
$$
where $\Delta_F$ is the Feynman propagator,
$$
   \Delta_F (t, \tau ) = { 1\over (2 \pi )^2 i}
   \int  \domega \dk { e^{i( \omega t - k \tau )} \over {
   \omega - k + i \varepsilon {\rm sgn} k}} \eqno\eq
$$

To solve for~$\alpha$ in terms of $\alpha_0$,
we expand~$\alpha$ in perturbation series
$$
   \alpha = \alpha_0+ g \alpha_1+g^2 \alpha_2 \cdots {\rm ,}
   \eqn\?
$$
and obtain
$$ \eqalign{ (\partial_t +\partial_\tau ) \alpha_0 & = 0 \cr
           ( \partial_t +\partial_\tau ) \alpha_1 & = -\partial_\tau \{
{ \alpha_0^2 \over 2\cosh^2 \tau } \} \cr
(\partial_t +\partial_\tau ) \alpha_2 & = -\partial_\tau \{
{ \alpha_0 \alpha_1  \over
\cosh^2 \tau } \} \cr
   (  \partial_t +\partial_\tau ) \alpha_3 &= -\partial_\tau \{ {1 \over
\cosh^2 \tau }
\left( \alpha_0 \alpha_2 + {1\over 2} \alpha_1^2 \right) \} \cr
    & \vdots \cr }  \eqn\per $$

Now the action can be expressed in terms of~$\alpha_0$. Taking the Fourier
transform,
the coeffecients of the expansion will be the desired $S$-matrix elements:
$$
   I[\alpha_0(w,k)] = \sum_0^\infty {2-n\over n!} S(\omega_1,k_1;
   \cdots \omega_n,k_n) \alpha_0(\omega_1,k_1)
   \alpha_0(\omega_2, k_2 ) \cdots \alpha_0( \omega_n,k_n )
   \eqno\eq
$$

If we calculate this
 perturbatively, using \per~, it is equivalent to the use of the ordinary
Feynman rules \refmark{\kr}.
However, we can solve for the $S$-matrix generating functional to all orders
if we know the exact solution of the non-linear classical equation.
A general parametric form was obtained by Polchinski\refmark{\polch}.
Introducing~$\sigma$ to
parameterize~$x$ and~$p$,
$$
   \eqalign{ x &= a( \sigma ) \sinh (t-\sigma) = \sinh \tau \cr
             p &=  a(\sigma) \cosh (t- \sigma )  \cr }
   \eqn \nonlin
$$
let us use a new field, $\epsilon (\sigma)$, to denote fluctuations
about the static background:
$$ a^2(\sigma) = 1- g \epsilon( \sigma) . $$

Now we shall express $\epsilon ( \sigma )$ in terms of
$\alpha_0  (x) $.
Since~$\epsilon(\sigma)$ may be thought of as encoding the
initial conditions, it will be natural and useful to define
in- and out-fields as follows:

$$ \eqalign{ \alpha_{\rm in \atop out} (t,\tau)
   & \equiv \alpha(t,\tau) - \int
   \dt' \dtau' \Delta_{R \atop A } (t-t ^ \prime , \tau - \tau' )
   ( \partial_{t'} + \partial _ { \tau ^ \prime} ) \alpha (t', \tau')
   \cr }, \eqn \inout
$$
where $\Delta_R$, $\Delta_A$ are {\it retarded} and {\it advanced }
propagators,
$$
   \Delta_{R \atop A} (t, \tau ) = { 1\over (2 \pi )^2 i}
    \int  \domega \dk { e^{i( \omega t - k \tau )} \over
   \omega - k \mp i \varepsilon } \eqno\eq
$$

We easily see that these fields match $\alpha$ in the distant
past and future:
$$ \eqalign{ \lim_{t \to -\infty} \alpha(t,\tau) & =  \lim_{t \to -\infty}
\alpha_{\rm in}(t,\tau) \cr
            \lim_{t \to \infty} \alpha(t,\tau) & =  \lim_{t \to
\infty}\alpha_{\rm out}(t,\tau),
    \cr } \eqn \han
$$
as follows from the property of \ra~
  We also see that the Fourier transforms of
$\alpha_{\rm in}$, $\alpha_{\rm out} $, and $ \alpha_0$
are related by
$$ \tilde \alpha_0(\omega,k) = \theta (k) \tilde\alpha_{\rm out}(\omega,k) +
\theta (-k)
\tilde\alpha_{\rm in}(\omega,k) \eqn\rel $$
which follows from \feyn~.
The relations~\han\ and~\rel\ uniquely determine~$\alpha$ in terms
of~$\alpha_0$, and any one of
$\alpha_0$,$\alpha_{\rm in}$ or $\alpha_{\rm out}$ contains all the information
about
the boundary value of the system.

Now we can use these results to express $ \epsilon (\sigma) $ in terms of $
\alpha_0 (\sigma) $.
 We consider the following situation.
Pick a point on the Fermi surface
with a given coordinate $x$ (or equivalently $\tau$), evolve the system
backwards in time to $ -\infty$, and then turn off the interaction;
then evolve forward back to the present time,
and call its resulting coordinate
$x_{\rm in}$ (or $\tau_{\rm in}$). Then we have the relation
$$ \eqalign{  x &= \sinh \tau = \sqrt{1-g \epsilon (\sigma) } \sinh (t- \sigma)
\cr
            x_{\rm in}  &= \sinh \tau_{\rm in} = \sinh (t- \tilde \sigma) \cr
            \tau(t= -\infty,\sigma) &= \tau_{\rm in}(t= -\infty,\tilde \sigma)
\cr}
\eqno \eq $$
{}From the last line of the above equation we get
 $$ \tilde \sigma = \sigma + {1 \over 2} \ln (1-g \epsilon (\ \sigma)) .\eqno
\eq  $$
Therefore
$$ \tau_{\rm in} (t,\tilde \sigma) = t- \tilde \sigma =  t - \sigma - {1 \over
2}
 \ln (1-g \epsilon (\sigma))  \eqno \eq $$
and we get
$$ \lim_{t \to -\infty} \alpha(\tau,t) = \lim_{t \to -\infty}
\alpha_{\rm in}(t - \tau) = \lim_{t \to -\infty} \alpha_{\rm in}(t - \tau_{\rm
in})  \eqn \imp
$$
In fact, for $t\to -\infty$,
$$
   \eqalign { \alpha &=
   {1 \over g } \cosh \tau [ \sqrt{ 1- g \epsilon (\sigma } )
   \cosh( t- \sigma ) - \cosh \tau ] \cr \cr
   &= { 1 \over g } \sqrt { (1-g \epsilon(\sigma) )\sinh^2 (t-\sigma) + 1 } \,
   \sqrt{ 1- g \epsilon ( \sigma) } \cosh ( t- \sigma ) \cr
   & \> - {1 \over g} \{ (1-g \epsilon(\sigma)) \sinh^2 \tau + 1  \} \cr \cr
    &= -  { \epsilon(\sigma) \over 2 } + {\cal O}(e^{2(t-\sigma)})  \cr }
   \eqno\eq
$$
Therefore,
$$ \epsilon (\sigma) = \epsilon_{\rm in}(\sigma + {1 \over2} \ln (1-
g\epsilon(\sigma))) ,  \eqn\income $$
and similarly,
$$ \epsilon (\sigma) = \epsilon_{\rm out}(\sigma - {1 \over2} \ln (1-
g\epsilon(\sigma))) ,\eqn\outgo $$
where  $\epsilon_{{\rm in} \atop {\rm out}} \equiv -2 \alpha_{{\rm in}\atop
{\rm out}} $.
We can invert these equations and get

$$ \epsilon_{{\rm in} \atop {\rm out}}(\sigma) = \epsilon(\sigma \mp {1 \over2}
\ln (1-
g\epsilon(\sigma))). $$

We can also derive from these equations the relation between $\epsilon_{\rm
in}$ and $\epsilon_{\rm
_out} $,

$$ \epsilon_{\rm in} (\sigma) = \epsilon_{\rm out}(\sigma - \ln (1-
g\epsilon_{\rm in}(\sigma))) ,\eqn\pol $$
which was previously derived in ref.~\refmark{\polch} and was used to calculate
the $S$-matrix. The details are given in Appendix A.

This type of equation, with  the left-hand side inside
the functional argument of the right-hand side,
was solved in ref.~\refmark{\moreeq}.
 Using that result, we get
$$
\tilde\epsilon_{{\rm in}\atop{\rm out}}(k) = -{1 \over 2 \pi g}
{1 \over 1 \mp  {i k / 2 }}
\int_{-\infty}^\infty \dsigma e^{-i k \sigma } [ (1- g \epsilon(\sigma))^{1 \mp
{i k /2} } - 1 ] \eqn\moore
$$
where $\tilde\epsilon$'s are the Fourier transforms.

Thus eqs.~$\moore$, \rel\ and \nonlin \
together determine the nonlinear field $p(t,x)$
in terms of $\alpha_0$ to all orders in $g$.

We next evaluate the action using this classical solution,
expanding in powers of the coupling~$g$.
$$
   \eqalign { I &= { 1 \over 2 \pi g^2 }  \int \dt \dsigma
({\partial x \over \partial \sigma } ) \left[ {1 \over 2} p(t,\sigma)
\partial_x^{-1} \dot p(t,\sigma) \,   + \, { p(t,\sigma)^3
\over 6 } \, - \, {1 \over 2 } ( x(t,\sigma)^2 + 1 ) p(t,\sigma) \right]
\cr \cr
                &= {1 \over 24 \pi g^2 }  \int \dt \int \dsigma ( w_0 + g w_1+
g^2 w_2 ) ,
\cr } \eqno \eq
$$
where
$$
\eqalign{ w_0 &= 4\cosh^4 (t-\sigma) , \cr
          w_1 &= [ -8 \cosh^4(t-\sigma) +3 \cosh^2(t-\sigma) ]
\epsilon(\sigma)
\cr
&\qquad   + [ 2 \cosh^3(t-\sigma) \sinh(t-\sigma) +  \cosh(t-\sigma)
\sinh(t-\sigma) ]\epsilon' , \cr
         w_2      & = [ 4\cosh^4(t-\sigma) - 3 \cosh^2 (t-\sigma) ]
\epsilon^2(\sigma)
\cr &\qquad
+ [ -2 \cosh^3(t-\sigma) \sinh(t-\sigma) +{3 \over 2 }
\cosh(t-\sigma) \sinh(t-\sigma) ] \epsilon \epsilon'. \cr
}
\eqno\eq
$$

To obtain the correct expansion in $g \epsilon$, we have to consider also the
contribution from the boundary.
Taking the range of spatial integral to be
$ \tau \equiv \sinh^{-1} x \in [-L,L]$,
(we take the limit $L \to \infty $ in the end),
$I[\epsilon(\sigma)]$ can be written in the following form.
$$ 24 \pi g^2 I = \int \dt \left[ \int_{\sigma_0(-L)+g \sigma_1(-L)+ \cdots}^
{\sigma_0(L)+g \sigma_1(L)+ \cdots} ( w_0 + g w_1+ g^2 w_2 ) \dsigma \right ]
\eqno
\eq $$
where $ \sigma(t,L) $ is given by
$$ ( 1-g \epsilon(t,L))^{1 \over 2} \sinh (t-\sigma(t,L)) = \sinh L \eqno \eq
$$
and
$$ \eqalign { \sigma(t,L) &= \sigma_0(t,L) + g \sigma_1(t,L) + \cdots \cr
                          &= t-L - {g \epsilon(\sigma) \over 2 } { \sinh L
\over \cosh L } + \cdots \cr } \eqno \eq $$

Therefore we have
$$ \eqalign { 24 \pi g^2 I &= \int \dt   \bigg[ \int_{t+L}^{t-L} ( w_0
+ g w_1 + g^2 w_2 )
\cr &
+ ( g\sigma_1 + g^2\sigma_2 + g^3 \sigma_3 + \cdots )( w_0
+ g w_1 + g^2 w_2 ) \vert^{\sigma = t-L}_{\sigma = t+L}  \cr
                 &  +     {1 \over 2!} ( g\sigma_1 + g^2\sigma_2 + g^3 \sigma_3
+ \cdots )^2 {\partial \over \partial \sigma} ( w_0 + g w_1 + g^2 w_2
)\vert^{\sigma = t-L}_{\sigma = t+L}
                   \cdots \cr
                  &  +      { 1 \over n!} ( g\sigma_1 + g^2\sigma_2 + g^3
\sigma_3 + \cdots )^n {\partial^{n-1} \over \partial \sigma^{n-1}}( w_0 + g w_1
+ g^2 w_2 )\vert^{\sigma = t-L}_{\sigma = t+L}
                   \cdots\bigg]
 \cr } \eqno \eq
$$

It is easy to check that $O(g\epsilon)$ term can be dropped and
$O((g\epsilon)^2)$ term vanishes. It is to be expected.
The $O(g\epsilon)$ term
drops out using the condition that $\int_{-\infty}^\infty dt \epsilon(t) = 0 $
since we are expanding around the solution
to the equation of motion with a fixed chemical potential.
The $O((g\epsilon)^2)$
term  can come only from the quadratic part of the Lagrangian $ \alpha (
 \partial_\tau^{-1}\partial_t + 1) \alpha $.
Once we expand the field $\alpha$ in
terms of linear field $\epsilon(t-\tau)$,
$$
   \alpha(t,\tau) = \epsilon_0(t-\tau) + g \epsilon_1(t,\tau) + \cdots
$$
the  $O((g\epsilon)^2)$ term is $\epsilon(t-\tau) \partial_\tau^{-1}(
\partial_t +
\partial_\tau ) \epsilon(t-\tau) = 0 $.
The $O(g^0)$ term is the static background which we drop from now on.

{}From the third order in $g\epsilon$, everything is boundary term in space
integration.

We have
$$  24 \pi g^2 I =
    \int_{-\infty}^{\infty} \dt \left [ {g^3 \epsilon^3 \over 4 } +{g^4
\epsilon^4 \over 8 } + {3 g^5 \epsilon^5 \over 40 } + { g^6 \epsilon^6 \over 20
} + \cdots \right ] \eqno \eq $$
We observe that any terms involving the derivatives comes in the form $
{\partial^m \over \partial t^m } ( \epsilon^n) $ and vanishes by $t$
integration. (We always begin with localized wavepacket and take the limit such
as making the length of the packet infinite after all the calculation is done.
Therefore, these kinds of surface terms vanish. )

This interesting result seems to hold at every order.
But unfortunately we could not prove it in general.
Here we make two comments.

\point For the sine-Gordon background which will be discussed in the next
section, Moore proved that $ \langle (e^{ipX+\beta_p})^n
(e^{ipX+\beta_p})^n \rangle $ is a $(2n-3)$rd polynomial in $p$ using the
conformal field theory calculation \refmark{\moresine}. (Although he could not
prove the explicit
form of it.) If any term containing derivatives contributed, this would take
down additional power of $p$ and the order of the polynomial would be higher
than $2n-3$. Therefore if one believes that the conformal field theory
calculation should give the same result as the collective field theory, then
the above statement is certainly true.

\point  The proof of scaling of sine-Gordon model for $p \to 0 $ limit
does not depend on the proof of the statement above. As will be clear when we
write down the explicit form of $\epsilon(\sigma)$ corresponding to the
sine-Gordon background, the derivative terms bring down non-zero powers of $p$
and these terms will vanish in $p \to 0 $ limit anyway.

\medskip

Once we realize that the derivative terms do not contribute, we can explicitly
integrate the lagrangian, treating $\epsilon$ as if it were a constant.

We have
 $$ \eqalign { I &= {1 \over 24 \pi g^2 }  \int_{-\infty}^{\infty}  \dt
\int_{-\sinh^{-1} ({ \sinh L \over \sqrt {1-g \epsilon}})}^{\sinh^{-1}({\sinh L
\over \sqrt {1-g \epsilon}})}  \dsigma [w_0+gw_1 + g^2 w_2 ] \cr
                 &= { 1 \over 12 \pi g^2 } \int \dt  \big[ -{3 \over 2} ( 1-
g\epsilon) \{ \log ( \sinh L + \sqrt{ \cosh^2 L -g \epsilon } ) - {1 \over 2}
\log (1-g \epsilon )
\}
\cr & \qquad \qquad \qquad
-{3 \over 2} \sqrt{ \cosh^2 L -g \epsilon } \sinh L    - ( \cosh^2 L - g
\epsilon )^{3 \over 2} \sinh L \big ] \cr
       & \cr
 & {\to \atop L\to \infty} {1 \over 16 \pi g^2} \int \dt   [ ( 1- g
\epsilon(t))
\log (1-g \epsilon(t)) + { g^2 \epsilon^2(t) \over 2}   ] + S_0   \cr
& \cr
&= { 1 \over 16 \pi g^2 } \sum_{n=3}^\infty \int \dt { g^n \epsilon^n(t) \over
n
(n-1) } + S_0 \cr
}  \eqn \fun $$
where $S_0$ is the static background and $O({1\over g})$ term is
dropped using the condition $ \int_{-\infty}^\infty dt \epsilon(t)=0$, as
mentioned before.

{\bf \chapter{ Gravitational sine-Gordon Model }}

The string theory in 1+1 dimensions with a nontrivial tachyon background
$T(\phi,X)$ is described by the worldsheet action

$$ \eqalign { S = & \int d^2 z \sqrt {\hat g} \left[ {1 \over 8 \pi} ( \hat
\nabla
\phi  )^2 + {\mu \over 16 \pi } e^{\sqrt 2 \phi } + { \sqrt 2 \over 4 \pi }
\phi R (\hat g ) \right] \cr
  &+ \int d^2 z \sqrt {\hat g}  \left[ {1 \over 4 \pi } ( \hat \nabla X )^2 +
T(\phi,X) \right] \cr } \eqn \world
$$
where $ \hat g $ is a fixed metric, $ \hat R $ is its curvature, $\mu$ is the
cosmological constant .
Expanding the partition function in $T(\phi,x)$, we get

 $$
Z \equiv \langle e^{-S} \rangle = \sum_{n=0}^\infty {1 \over n! } \langle
e^{ik_1 X+\beta_{k_1} \phi}
  \cdots e^{ik_n X+\beta_{k_n} \phi} \rangle_{T=0}
\tilde T(k_1) \tilde T(k_2) \cdots \tilde T(k_n) \eqno \eq
$$
where $\langle$ \ $ \rangle$ indicates the path integration over $\phi$,
$X$, and $\tilde T$ is  defined by
$$
   T(x) = \int \dk \,\tilde T(k) e^{ikX+\beta_k  \phi }
   \eqn\?
$$
and $\beta_p = ( \sqrt 2 - {\vert p \vert \over 2 } )$.

 Since $e^{ipX+ \beta_p \phi } $ is the tachyon vertex operator, the
expansion coefficients $ \langle e^{ik_1 X+\beta_{k_1} \phi} \cdots
\rangle_{T=0} $
are nothing but the tachyon scattering amplitudes on a flat background.
Therefore by identifying $T(k)$ with $\alpha_0(k)$ in collective
field  theory, we see that the worldsheet action $S$ given by \world\ would
correspond to the $S$-matrix generating functional in that theory.

The gravitational sine-Gordon model is described by the action
\world \ when $T(X) = \lambda e^{\beta_p \phi } \cos(pX)$.
We can apply the formalism we developed in the previous chapters to
calculate this action and get the scaling behavior.
The physics of this model
is discussed in \refmark{\moresine,\hsu,\sch}.
It was conjectured in ref.~\refmark{\moresine} that
$$
\langle T_p^n T_{-p}^n \rangle
= - \mu^{np-2n+2} n! p^{2n}(1-p)^n
{\Gamma( n(1-p)+n-2) \over \Gamma( n(1-p)+1) },
$$
where these scattering amplitudes are calculated at the tree level.
If this formula is correct,
then the partition function of sine-Gordon model on
the sphere is
$$
   \eqalign{ {\cal Z}_0(\mu, \lambda)
           & = \sum_{n=0}^\infty \langle ( \cos (pX) e^{\beta_p
\phi} )^n \rangle_{\lambda =0} \cr
                      & = \sum_{n=0}^\infty { ({1 \over 4} \lambda^2 )^n
\over (n!)^2 } \langle (e^{ipX+ \beta_p \phi} )^n (e^{-ipX+ \beta_p \phi})^n
\rangle_{\lambda = 0 } \cr
                       &= -\sum {1 \over n!} \mu^{np-2n+2} p^{2n}(1-p)^n
{\Gamma( n(1-p)+n-2) \over \Gamma( n(1-p)+1) } \cr
&= \mu^pp - {\mu^{2p-2} \over 2} p^4 (1-p) - \mu^{3p-4}p^6 (1-p)^3+ \cdots \cr
} \eqn\mooresine $$
Here $\mu$ corresponds to ${1 / g}$ of previous chapters. The action above
is convergent for $p<2$ and scales like $(g-g_c)^{5 / 2}$ (or equivalently,
$(\mu-\mu_c)^{5/2}$), which
is like $c=0$ theory.
It is argued in refs.~\refmark{\hsu, \moresine,  \sch} that there is a
phase transition between $c=0$ and $c=1$ model at $p=2$. This situation is
similar to that of discrete chain of matrices
\Ref\pari{G.Parisi \journal Phys.Lett. & 238B (1990) 213; \nextline D.J.Gross
and  I.R.Klebanov \journal Nucl.Phys. & B344 (1990) 475 }.
To see the scaling most easily, take the derivative of \mooresine
\ three times with respect to $\mu$ and see whether it diverges like
$(\mu-\mu_c)^{- 1 / 2 }$.
 First, consider the behavior of $$(\mu^s-\mu_c^s)^\gamma = \sum_{n=0}^\infty {
\gamma! \over n! (n-\gamma)! }(-{\mu \over \mu_c})^{sn} \mu_c^s\gamma  \eqn
\series $$
for some $\gamma < 0 $. As $ \mu \to \mu_c $, the above expression
diverges, so the most of the contribution comes from the large~$n$ terms in the
series. Therefore it is enough to look at the large~$n$ behavior of the
coefficients of the above expansion to get the critical exponent~$\gamma$.
Using the Stirling's formula, we get
$$ \eqalign{ \log { \gamma! \over n! (n-\gamma)! } &=  \log [(-1)^n
{(n-\gamma-1)! \over n! (\gamma-1)! }] \cr
                        &= i n \pi - ( \gamma + 1) \log n - \log (-\gamma-1)! +
O(1/n) \cr}.\eqn \first $$
Therefore
$$ (\mu^s - \mu_c^s)^\gamma \sim  \sum n^{-(\gamma+ 1)} ({\mu \over \mu_c
})^{sn} \eqn \second $$
up to a constant normalization.

Now taking the derivatives of \mooresine \ three times with respect to $\mu$,
we get
$$ \eqalign{
   {\partial^3 {\cal Z}_0(\mu, \lambda) \over \partial \mu^3 }
         &= \sum {1 \over n!} \mu^{np-2n-1} p^{2n}(1-p)^n {\Gamma( n(2-p)-1)
\over \Gamma( n(1-p)+1) } \cr }
\eqno \eq .
$$
Using Stirling's formula again, we get
$$ {\partial^3 {\cal Z}_0(\mu, \lambda) \over \partial \mu^3 } \sim \sum \left
[{
\lambda^2 (2-p)^{2-p} \mu^{p-2} p^2 \over (1-p)^{-p}} \right ]^n{1 \over \sqrt
n }
\eqn\fourth . $$

Comparing  \fourth with \second, we see that $$\gamma = -{1 \over 2}$$
and
$$
\mu_c(p)={(2-p) \over (1-p)} (\lambda \vert 1-p \vert )^{ 2 / (2-p) }
 \eqn\crit
$$
as discussed in ref.~\refmark \hsu.

  Let us reproduce this result using the collective field method.
According to
what we have learned in previous chapters, we expect to get this result up to
normalizations of the field if we
use
$$ \tilde\epsilon_0(k) = \alpha \delta(k+p) + \beta \delta(k-p) {\rm ,} $$
and analytically continue $p \to ip $  in the end, since we are doing
calculation in Minkowski space-time.   Note here we are now extending  the
field $\alpha$ to take complex values.  We get from \rel,
$$ \tilde \epsilon_{\rm in} (k) =  f(k) \theta(k) +  \alpha \delta(k+p)
\eqn\sinin
$$
$$ \tilde \epsilon_{\rm out} (k) =  f(k) \theta(-k) +  \beta \delta(k-p)
\eqn\sinout $$
where $f(k)$ and $\epsilon(\sigma)$ are the
two functions to be determined from
two equations \moore,
$$  f(k) \theta(k) + \alpha \delta(k+p) = -{1 \over 2 \pi g} {1 \over 1 -  i k
/
 2 }
\int_{-\infty}^\infty \dsigma e^{-i k \sigma } [ (1- g \epsilon(\sigma))^{1 -
i k / 2 } - 1 ] \eqn\sinmoor $$
$$  f(k) \theta(k) + \beta \delta(k-p) = -{1 \over 2 \pi g} {1 \over 1 +  i k
/ 2 }
\int_{-\infty}^\infty \dsigma e^{-i k \sigma } [ (1- g \epsilon(\sigma))^{1 +
i k / 2 } - 1 ] \eqn\sinmore $$
  From $\theta (-k) \times$ \sinmoor and $\theta (k) \times$ \sinmore we get
the decoupled equations
$$ \alpha \delta(k+p) =  -{\theta(-k) \over 2 \pi g} {1 \over 1 -  i k / 2
}
\int_{-\infty}^\infty \dsigma e^{-i k \sigma } [ (1- g \epsilon(\sigma))^{1 -
i k / 2 } - 1 ] \eqn\nega $$
$$ \beta \delta(k-p)= -{\theta(k) \over 2 \pi g} {1 \over 1 +   i k / 2 }
\int_{-\infty}^\infty \dsigma e^{-i k \sigma } [ (1- g \epsilon(\sigma))^{1 +
i k / 2 } - 1 ] \eqn\posi $$
 These equations determine $\epsilon(\sigma)$ to all orders uniquely for the
sine-Gordon solution. We get
$$ \eqalign{ \epsilon_0 (\sigma) &= \alpha e^{-ip \sigma} + \beta e^{ip \sigma}
\cr
 \epsilon_1 (\sigma) &= {ip \over 2} (\alpha^2 e^{-2ip \sigma} + \beta^2 e^{2ip
\sigma}) \cr
 \epsilon_2 (\sigma) &= (-{3 p^2 \over 8} + {ip \over 4} ) (\alpha^3 e^{-3ip
\sigma} + \beta^3 e^{3ip \sigma}) \cr
                     &+ (-{p^2 \over 8} + {ip \over 4} ) (\alpha^2 \beta
e^{-ip\sigma} + \alpha \beta^2 e^{ip \sigma} ) \cr
                     & \vdots \cr . } \eqno\eq  $$
( The list of $\epsilon_n$ up to $n = 8$ is given in Appendix B. )
Plugging these into \fun, we get
$$ \eqalign{ I_4 &= { 1 \over 16 \pi} ( ip+1)  \cr
                I_6 &= {1 \over 12 \pi } ( ip+1)^3  \cr
             I_8 &= {1 \over 32 \pi} ( ip+1) (4 ip + 5)  \cr
              I_{10} &= {1 \over 120 \pi } ( ip+1)^5 (5ip + 6 ) (5ip + 7)  \cr
&\vdots \cr
} \eqno \eq $$
where $I_n$ is the coefficient of $g^{n-2}$ in the expansion of $I$,
which agrees with the formula
$$ I = \sum_{n=2}^\infty
{1 \over 4 \pi g^2 } { (n-1) \over n! } { \{ n(2+ip)-3
\} ! \over \{ n (1+ip) \} ! } (1+ip)^n \eqno\eq $$
This formula can be proved at least in the limit $p \to 0 $

It is easy to prove that $\epsilon_n(\sigma)$ ($n>1$) generated
perturbatively by \nega \ and \posi \  always has nonzero powers of $p$
as a prefactor. (We can prove this by mathematical induction, for example)
therefore when we plug in $\epsilon$ into the action and and take $p \to 0$
limit, all the contributions from   $\epsilon_n(\sigma)$'s ($n>1$) will vanish.
Therefore we may take
$$ \epsilon(\sigma) = \epsilon_0(\sigma) = \alpha e^{-ip\sigma} + \beta
e^{ip\sigma} $$
Note that we cannot put $p$ in $e^{ip\sigma}$ to $0$ since we have to do the
$ \sigma$
integral first. This is the difference from $p=0$ case which is just a shift of
cosmological constant in $c=1$  theory.
We have
$$ \eqalign{ 16\pi g^2 I &= \sum_{n=3}^\infty \int dt { g^n \epsilon^n(t) \over
n (n-1) } \cr
                         &= \sum_{n=2}^\infty { 1 \over 2n(2n-1) } { (2n)!
\over (n!)^2 } g^{2n} \alpha^n \beta^n T \cr
                         &= 2 \sum_{n=2}^\infty {(2n-3)! \over (n!)^2 } (n-1)
(\alpha \beta )^n g^{2n}T} \eqno\eq $$
where $T$ is the volume of time. This proves the Moore's conjecture for $p \to
0
 $ limit and therefore shows
the correct scaling behavior.

{\bf \chapter{ Conclusion }}

In this paper we developed a formalism  in collective field theory with which
we can describe a general  time-dependent tachyon background at the tree level.
 The background is specified by a time-dependent classical solution, which in
turn  is decided by the initial data, the linearized field.  We applied this
formalism to the gravitational sine-Gordon model.
We reproduced the calculations of refs.~\refmark{\hsu,\ \moresine}; \ie
we could perturbatively check
the conjecture for the scaling.
We went further to prove the conjecture for
the~$p \to 0$ limit.

It was claimed in ref.~\refmark{\hsu}
that the model corresponds to many copies of the $c=0$ model.
Since the present calculation is at tree level, this could not be seen.
Also for general $p$, although we have an equation
specifying the background to all orders in~$g$, we could not find a simple
argument which shows the interesting properties such as phase transition.
These points may need further study.

\ack

I give special thanks to Antal Jevicki, who gave me many helpful advices, and
to Paul Mende, for useful discussions and careful editing of this
paper.

\Appendix{A: Properties of the equations of the type \pol \ }

 We consider an equation of the type

$$ f(x) = g(x-a\ln (1-bf(x)) $$

First we show how this can be inverted.
We consider a neighborhood of a given point $x$ where the $f^{-1}$ is well
defined and put

$$ x=f^{-1}(y) {\rm .} $$

Then it can be written

$$ y= g(f^{-1}(y) -a \ln(1-by)) $$

Inverting $g$, we get

$$ g^{-1}(y) = f^{-1}(y) -a \ln(1-by) $$

which is

$$ f^{-1}(y) = g^{-1}(y)+ a\ln (1-by) $$

Putting it inside the argument of $f$, we get

$$ y= f(g^{-1}(y)+a\ln(1-by)) $$

Finally we substitute $g(x)$ for $y$ to get

$$ g(x)= f(x+a\ln(1-bg(x))) $$

Next, I we consider the two equations

$$ h(x)=f(x-a\ln(1-bh(x)))  $$

$$ h(x)=g(x+c\ln(1-dh(x)))  $$

and eliminate $h(x)$ to get the relation between $f(x)$ and $g(x)$ .

Again, putting $h(x)=y$, we get

$$ \eqalign { f^{-1}(y)&= x-a\ln(1-by) \cr
              g^{-1}(y)&= x+c\ln(1-dy) \cr } $$

 Subtracting these two, we  get

$$ f^{-1}(y) = g^{-1}(y)-a\ln(1-by)-c\ln(1-dy) $$

Finally we put this inside the argument of $f$ and put $y=g(x)$ to get

$$ g(x) = f(x - a\ln(1-bg(x))-c\ln(1-dg(x))) $$

This argument proves the equivalence of \income \ and \outgo \ to \pol \ where
in this case $b=d, a=c={1 \over2} , g(x)= \epsilon_{\rm in}, f(x)=
\epsilon_{\rm
 out } $  .

\vfill

\Appendix{ B: The list of $\epsilon_n$ up to $n=8$ }

$$ \eqalign{ \epsilon_0 (\sigma) &= \alpha e^{-ip \sigma} + \beta e^{ip \sigma}
\cr
\cr
\epsilon_1 (\sigma) &= {ip \over 2} (\alpha^2 e^{-2ip \sigma} + \beta^2 e^{2ip
\sigma}) \cr
\cr
  \epsilon_2 (\sigma) &= (-{3 p^2 \over 8} + {ip \over 4} ) (\alpha^3 e^{-3ip
\sigma} + \beta^3 e^{3ip \sigma}) \cr
                     &+ (-{p^2 \over 8} + {ip \over 4} ) (\alpha^2 \beta
e^{-ip\sigma} + \alpha \beta^2 e^{ip \sigma} ) \cr
\cr
  \epsilon_3 (\sigma) &= (-{i p^3 \over 3} - {p^2 \over 2} + {ip \over
6} ) (\alpha^4 e^{-4ip
\sigma} + \beta^4 e^{4ip \sigma}) \cr
& +(-{i p^3 \over 6} -{ p^2 \over 2 } + {ip \over 3 } ) (\alpha^3 \beta e^{-2ip
\sigma} + \beta^3 \alpha  e^{2ip \sigma}) \cr
\cr
 \epsilon_4 (\sigma) &= ({125 p^5 \over 384} - { 25 i p^3 \over 32 }
 - { 55 p^2 \over 96} + {ip \over 8} ) (\alpha^5 e^{-5ip
\sigma} + \beta^5 e^{5ip \sigma}) \cr
& +({27 p^4 \over 128 } -{ 27 i p^3 \over 32} - {33 p^2 \over 32}
+ {3ip \over 8} ) (\alpha^4 \beta e^{-3ip
\sigma} + \beta^4 \alpha e^{3ip \sigma}) \cr
& +({17 p^4 \over 192 } -{3 i p^3 \over 8} - { 25 p^2 \over 48} + {ip \over
4} ) (\alpha^3 \beta^2 e^{-ip
\sigma} + \beta^3 \alpha^2 e^{ip \sigma}) \cr
\cr
 \epsilon_5 (\sigma) &=({ 27 i p^5 \over 80 } + {9p^4 \over 8 } -{21
 i p^3 \over 16} - {5p^2 \over 8} + {ip \over 10} ) (\alpha^6 e^{-6ip
\sigma} + \beta^6 e^{6ip \sigma}) \cr
&+({ 4 i p^5 \over 15 } + {4 p^4 \over 3 } -{7
 i p^3 \over 3} - {5p^2 \over 3} + {2ip \over 5} ) (\alpha^5 \beta
 e^{-4ip
\sigma} + \beta^5 \alpha e^{4ip \sigma}) \cr
&+({7 i p^5 \over 48 } + {19p^4 \over 24 } -{77
 i p^3 \over 48} - {35p^2 \over 24} + {ip \over 2} ) (\alpha^4 \beta ^2 e^{-2ip
\sigma} + \beta^4 \alpha^2 e^{2ip \sigma})\cr
\cr }  $$

 $$
\eqalign{
 \epsilon_6 (\sigma) &= (-{16807 p^6 \over 46080 } + { 2401 i p^5 \over 1536 }
+ {5831 p^4 \over 2304 } -{245
 i p^3 \over 128} - {959 p^2 \over 1440 } + {ip \over 12} ) (\alpha^7 e^{-7ip
\sigma} + \beta^7 e^{7ip \sigma}) \cr
&-({3125 p^6 \over 9216 } - { 3125 i p^5 \over 1536 } - {10625 p^4 \over 2304 }
+ {625
 i p^3 \over 128} + {685 p^2 \over 288 } -  {5ip \over 12} ) (\alpha^6 \beta
e^{-5ip
\sigma} + \beta^6 \alpha e^{5ip \sigma}) \cr
&-({1107 p^6 \over 5120 } - { 729 i p^5 \over 512 } - {951 p^4 \over 256 }
+{615  i p^3 \over 128} +{489 p^2 \over 160 } - {3ip \over 4} ) (\alpha^5
\beta^2 e^{-3ip
\sigma} + \beta^5 \alpha^2 e^{3ip \sigma}) \cr
&- ({929 p^6 \over 9216 } - { 341 i p^5 \over 512 } -{4025 p^4 \over 2304 }
+{881
 i p^3 \over 384} + {439 p^2 \over 288 } - {5ip \over 12} )
 (\alpha^4 \beta^3 e^{-ip
\sigma} + \beta^4 \alpha^3 e^{ip \sigma}) \cr
\cr
\epsilon_7 (\sigma) &=-({128 \over 315} p^7 + {32 p^6 \over 15 } - { 40 i p^5
\over 9 } - {14 p^4 \over 3 } +{116
 i p^3 \over 45} +{7 p^2 \over 10 } - {ip \over 14} ) (\alpha^8
 e^{-8ip
\sigma} + \beta^8  e^{8ip \sigma}) \cr
&-({243 \over 560} p^7 + {243 p^6 \over 80 } - { 135 i p^5 \over 16 } - {189
p^4 \over 16 } +{87
 i p^3 \over 10} + {63 p^2 \over 20 } - {3ip \over 7} ) \cr
& \qquad \times (\alpha^7
 \beta e^{-6ip
\sigma} + \beta^7 \alpha  e^{6ip \sigma}) \cr
&-({14 \over 45} p^7 + {12 p^6 \over 5 } - { 137 i p^5 \over 18 } - {38 p^4
\over 3 } +{521
 i p^3 \over 45} + {163 p^2 \over 30 } - ip  ) \cr
 & \qquad \times (\alpha^6
 \beta^2 e^{-4ip
\sigma} + \beta^6 \alpha^2  e^{4ip \sigma}) \cr
&-({131 \over 720} p^7 + {343 p^6 \over 240 } - { 673 i p^5 \over 144 } - {131
p^4 \over 16 } +{733
 i p^3 \over 90} + {263 p^2 \over 60 } - ip  ) \cr
& \qquad \times (\alpha^5
 \beta^3 e^{-2ip
\sigma} + \beta^5 \alpha^3  e^{2ip \sigma})
\cr
\cr
\epsilon_8 (\sigma) &=-({ 531441 p^8 \over 1146880} - {59049 p^7 \over 20480}
- {150903 p^6 \over 20480 } + { 5103 i p^5 \over 512 } \cr
& + {78327 p^4 \over 10240 } -{4221
 i p^3 \over 1280} - {3267 p^2 \over 4480 } + { ip \over 16}  ) \cr
& \qquad \times (\alpha^9 e^{-9ip
\sigma} + \beta^9  e^{9ip \sigma}) \cr
&+({ 823543 p^8 \over 1474560 } - {823543 p^7 \over 184320 } - {2705927 p^6
\over 184320 } + { 117649 i p^5 \over 4608 } \cr
& + {2321767 p^4 \over 92160 } -{160867
 i p^3 \over 11520} - {2541 p^2 \over 640 } + {7ip \over 16 }  ) \cr
 & \qquad \times  (\alpha^8
 \beta e^{-7ip
\sigma} + \beta^7 \alpha  e^{7ip \sigma}) \cr
&+({ 228125 p^8 \over 516096 } - {71875 p^7 \over 18432 } - {66875 p^6 \over
4608 } + { 135625 i p^5 \over 4608 } \cr
& + {160805 p^4 \over 4608 } -{6895
 i p^3 \over 288} - {5825 p^2 \over 672 } + {5ip \over 4 }  ) \cr
 & \qquad \times (\alpha^7
 \beta^2 e^{-5ip
\sigma} + \beta^7 \alpha^2  e^{5ip \sigma}) \cr
} $$
$$ \eqalign{
&-({ 11907 p^8 \over 40960 } - {27009 p^7 \over 10240 } - {26199 p^6 \over 2560
} + { 11293 i p^5 \over 512 }\cr
& + {73093 p^4 \over 2560 } -{7103
 i p^3 \over 320} - {1531 p^2 \over 160 } + {7ip \over 4 }  ) \cr
 & \qquad \times (\alpha^6
 \beta^3 e^{-3ip
\sigma} + \beta^6 \alpha^3  e^{3ip \sigma}) \cr
&+({ 106049 p^8 \over 737280 } - {29933 p^7 \over 23040 } - {460909 p^6 \over
92160 } + { 24581 i p^5 \over 2304 } \cr
& + {629921 p^4 \over 46080 } -{30439
 i p^3 \over 2880} - {4411 p^2 \over 960 } + {7ip \over 8 }  ) \cr
 & \qquad \times (\alpha^5
 \beta^4 e^{-ip
\sigma} + \beta^5 \alpha^4  e^{ip \sigma}) \cr
             . }    $$

\vfill

\refout

 \end